\documentclass{JINST}
\usepackage{overpic}
\usepackage{amsmath,amssymb,wasysym}
\providecommand*{\un}[1]{\ensuremath{\mathrm{~#1}}}

\title{Characterization of bolometric Light Detectors for rare event searches}

\author{
J.W.~Beeman$^a$, F.~Bellini$^{b,c}$, N.~Casali$^{d,e}$, L.~Cardani$^{b,c}$, I.~Dafinei$^c$, S.~Di~Domizio$^{f,g}$, F.~Ferroni$^{b,c}$, L.~Gironi$^{h,i}$, S.~Nagorny$^e$, F.~Orio$^c$\thanks{Corresponding author.}, L.~Pattavina$^e$, G.~Pessina$^i$, G.~Piperno$^{b,c}$, S.~Pirro$^i$, E.~Previtali$^i$, C.~Rusconi$^i$, C.~Tomei$^c$, M.~Vignati$^c$\\
\llap{$^a$}Lawrence Berkeley National Laboratory , Berkeley, California 94720, USA\\
\llap{$^b$}Dipartimento di Fisica - Universit\`{a} di Roma La Sapienza, I-00185 Roma - Italy\\
\llap{$^c$}INFN - Sezione di Roma, I-00185 Roma - Italy\\
\llap{$^d$}Dipartimento di Scienze Fisiche e Chimiche - Universit\`{a} degli studi dell'Aquila, I-67100 Coppito (AQ) - Italy\\
\llap{$^e$}INFN - Laboratori Nazionali del Gran Sasso, I-67010 Assergi (AQ) - Italy\\
\llap{$^f$}Dipartimento di Fisica, Universit\`{a} di Genova, I-16146 Genova - Italy\\
\llap{$^g$}INFN - Sezione di Genova, I-16146 Genova - Italy\\
\llap{$^h$}Dipartimento di Fisica - Universit\`{a} di Milano Bicocca, I-20126 Milano - Italy\\
\llap{$^i$}INFN - Sezione di Milano Bicocca, I-20126 Milano - Italy\\
\llap{$^*$}e-mail: \email{filippo.orio@roma1.infn.it}
}

\abstract{
Bolometers have proven to be very good detectors to search for rare processes thanks to their excellent energy resolution and their low intrinsic background. 
Further active background rejection can be obtained by the simultaneous readout of the heat and light signals produced by particles interacting in scintillating bolometers, as proposed by the LUCIFER experiment. 
In this framework, the choice of the light detector and the optimization of its working conditions play a crucial role.
In this paper, we report a study of the performances of a Germanium bolometric light detector in terms of signal amplitude, energy resolution and signal time development. The impact of various operational parameters on the detector performances is discussed.
}
\keywords{Detectors of Radiation, Scintillating bolometers, Rare processes}

\begin{document}       
\section{Introduction}
Bolometers \cite{bolo-review1,bolo-review2} have proven to be excellent detectors for application to the search for rare events, like neutrinoless double beta decay and dark matter interactions. This is mainly due to their good energy resolution and to the wide choice offered in terms of the absorber material. Using bolometers, it is often possible to match the characteristics of the detector material with the requirements of rare event searches.

The strongest requirement, when coming to neutrinoless double beta decay (${\rm 0\nu\beta\beta}$), is the reduction of the radioactive background coming both from environmental sources and from contaminations in the detectors and the surrounding materials~\cite{Avignone:2007fu}. In particular, $\alpha$ background assumes a crucial role~\cite{Clemenza:2011zz}. The expected ${\rm 0\nu\beta\beta}$ signal consists of two simultaneous electrons with a summed energy of a few MeV, an energy which can be released also by degraded $\alpha$ particles emitted from radioactive contaminations on the surface of the detectors or of the facing materials.
The objective of the future experiments is challenging: the background level in the region of interest for ${\rm 0\nu\beta\beta}$ must be below $10^{-3}\un{counts/(kg~keV~y)}$. This cannot be achieved with standard bolometers, due to the lack of active $\alpha$ background rejection methods. 

Bolometers are sensitive calorimeters operated at $\sim 10\un{mK}$ where the detected signal is the temperature rise produced by the energy deposited in particle interactions. The various kind of particles ($\beta$/$\gamma$s, $\alpha$s, neutrons and recoiling nuclei) interact in different ways inside the absorber material but, in the hypothesis that any kind of energy deposition is eventually converted into heat, the heat signal recorded in the bolometer shows only a slight dependence on the particle type. 

Particle identification in bolometers can be obtained by the simultaneous readout of the heat and the light signal produced by scintillation in scintillating materials~\cite{de2003experimental,cozzini2004detection,Arnaboldi:2010tt,ZnSe2013,Beeman:2012gg,Beeman:2013Pb} or by ${\rm \check{C}}$erenkov effect in non-scintillating crystals~\cite{Beeman:2011yc}.
Indeed, it has been observed in both cases that the amount of emitted light has a strong dependence on the particle type.

The light detectors used so far for scintillating bolometers and for ${\rm \check{C}}$erenkov light readout consist in thin germanium slabs, also operated as bolometers. 
This choice was driven by several considerations. Being opaque semiconductors, they are sensitive over an extremely wide range of photon wavelengths and satisfy the very stringent radiopurity requirements of rare event searches. Their overall quantum efficiency can be as good as the one of photodiodes, providing at the same time an higher energy resolution. Moreover, they are much easier to operate at cryogenic temperatures. 

The operational characteristics of the light detectors (response function, energy resolution, noise, etc...) are crucial for the optimization of their overall performances, and consequently for the improvement of the background rejection capability.

In this paper we present the results of the characterization of a light detector used for bolometric applications. The results were obtained in low temperature experimental runs carried out  at Laboratori Nazionali del Gran Sasso (LNGS), near L'Aquila - Italy.
 
\section{Light detector operation}\label{sec:ld_operation}
 
We operated as bolometer a light detector (LD) produced by UMICORE, constituted by a disk-shaped pure Ge crystal ($\diameter 50\un{mm} \times 300\un{\mu m}$) grown using Czochralski technique. One side of the wafer was etched, the other one was polished. 

The temperature sensor of the LD is a $3 \times 1.5 \times 0.4\un{mm^3}$ Neutron Transmutation Doped (NTD) Ge thermistor~\cite{Itoh:1996}, thermally coupled to the etched side of the crystal via 6 epoxidic glue spots of $\approx 600\un{\mu m}$ diameter and $\approx 50\un{\mu m}$ height. 

The scintillation photons, as well as other particles, interact with the LD and deposit their energy within its volume. The energy deposition is eventually converted into phonons, causing a temperature rise of the system and, in particular, of the NTD thermistor. 

The thermistor works as a resistive device that converts temperature variations into resistance variations. Its resistance can be expressed as:

\begin{equation}
R(T) = R_0\,{\rm exp}\,\left (\frac{T_0}{T}\right )^\gamma
\label{eq:NTDres}
\end{equation}
where $R_0, T_0$ and $\gamma$ depend on the doping level and are determined experimentally (typical values are: $R_0 = 1.15\,\Omega$, $T_0 = 3.35$ K and $\gamma = 1/2$).
 
\begin{figure}[t]
\centering \includegraphics[width=0.4\textwidth]{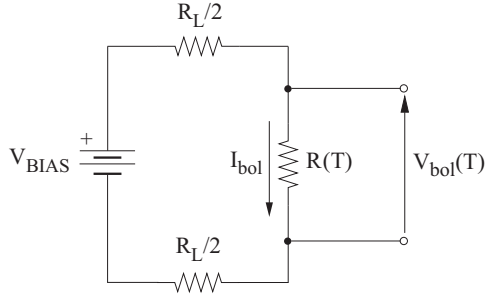}
\caption{Electric scheme of the bias circuit used for thermistor readout.}
\label{fig:bias_circuit}
\end{figure}

To measure the resistance variation, a bias voltage $V_{\rm BIAS}$ is applied across a pair of load resistors $R_L / 2$ in series with the thermistor, as shown in Fig.~\ref{fig:bias_circuit}.
The load resistance $R_L$ is chosen much higher than the thermistor resistance $R(T)$ for noise reasons. The resulting current in the circuit $I_{bol}$ remains almost constant when the temperature changes. 
The steady current produces a power dissipation which increases the temperature of the sensor and acts back on its resistance $R(T)$, modifying its value according to eq.~\ref{eq:NTDres}. 
Any energy release in the wafer causes a further temperature rise. Since the thermistor is thermally coupled to the LD,  this will result in a variation of $R(T)$ and, therefore, of the voltage across the thermistor $V_{bol}$, leading to a signal.
Load resistances used in this work range from $2\un{G\Omega}$ to $11\un{G\Omega}$, while the bias voltage was varied between 0.5 and $30\un{V}$.

The signals are amplified by means of Silicon JFETs and fed into an 18 bit NI-6284 PXI ADC unit~\cite{Arnaboldi:2006mx,Arnaboldi:2004jj}. The ADC is preceded by an analogic 6-pole low-pass Bessel-Thomson filter, with the roll-off rate of $120\un{dB/decade}$, to prevent aliasing effects on the acquired signal. 
The trigger is software generated on each bolometer and, when it fires, waveforms $250\un{ms}$ long, sampled at $8\un{kHz}$, are saved on disk. To maximize the signal to noise ratio, waveforms are processed offline with the optimum filter algorithm~\cite{Gatti:1986cw,Radeka:1966}.

\subsection{Noise contributions}
In a scintillating bolometer, only a few percentage of the particle energy is converted into scintillation photons. As a consequence, the corresponding light signals produced in the LD have energies of the order of few tens of keV~\cite{Beeman:2012gg,Arnaboldi:2010jx}. Dealing with small signals, it is fundamental to identify the different noise contributions affecting the measurements.

The dominant noise contribution arises from mechanical vibrations of the cryogenic apparatus, propagated to LD and wires. Vibrations cause temperature fluctuations in the volume of the LD. These instabilities are a very dangerous source of noise since they have a frequency spectrum similar to the signal one, that is within the range $0-100 \un{Hz}$. In addition, the wire vibrations are a source of microphonic noise. The thermodynamic fluctuations of the light detector give a negligible contribution to the noise compared to the aforementioned sources.

Other noise contributions come from electronics. The first one is the parallel Johnson noise of the load resistors, that develops across the dynamic impedance of the thermistor. As a consequence, its effect is larger at high values and becomes negligible at small values of the thermistor impedance. Considering the temperature of the resistors (room temperature) and the typical parameters of our setup, the resulting contribution of this kind of noise is expected to be of the order of $1 - 30\un{eV}$ rms. 
Further noise contributions are given by the preamplifiers and consist in a series white noise ($30 - 120\un{eV}$ rms) generated by the JFET resistances, and  a series $1/f$ noise ($5 - 25\un{eV}$ rms). Being independent from thermistor impedance, these noises have negligible effects at large values and become dominant at small values of the thermistor impedance.
The ADC quantization noise contribution was found to be minimal. We must remember that all the cited sources must be quadratically summed.

\section{Experimental setup} 

The characterization of the LD was structured in a set of measurements carried out in an Oxford 200 $^3$He/$^4$He dilution refrigerator located deep underground in the Laboratori Nazionali del Gran Sasso. 
The working temperature of the cryostat was varied between 10 and $20\un{mK}$. 

\begin{figure}[htb]
\centering \includegraphics[width=0.3\textwidth]{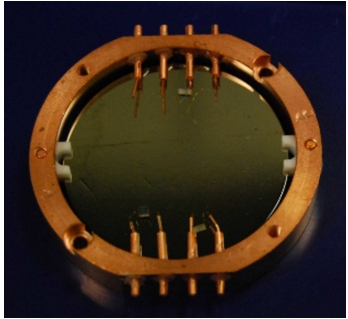}
\caption{Mounting of the LD into the copper frame.}
\label{fig:setup}
\end{figure}

The LD was mounted in a copper frame and held in place by PTFE clamps, as shown in Fig.~\ref{fig:setup}. The frame was accommodated in a container, also made of copper, connected to the refrigerator by means of a spring to reduce vibrations.

To allow a proper calibration of the LD signal, a $^{55}$Fe source, producing two X-rays at 5.9 and $6.5\un{keV}$, was faced to the LD.
The choice of such low energy calibration lines is due to the fact that, as previously mentioned, light signals produced in scintillating bolometers have typical energies of the order of $10\un{keV}$. 

\section{Detector performances}
We collected about 500 hours of data. The data-taking was split in short-lasting runs of few hours. 
As the main goal of the measurements was to characterize the LD and to optimize its performances, during these runs we systematically changed the working conditions, varying the bias voltage, the load resistance, the Bessel-Thomson cutoff frequency and the temperature.
In the following sections we present the obtained results, trying to underline the effect of each operational parameter on the LD performances.

\subsection{Variation of the bias current} \label{sec:bias}

\begin{figure}[t]
\centering \includegraphics[width=0.48\textwidth]{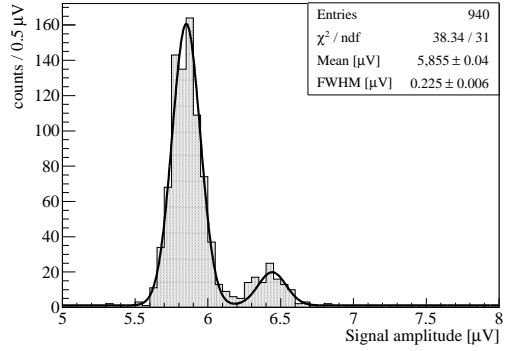}
\caption{Fit of the $^{55}$Fe X-ray double peak, obtained by biasing the LD with a current $I_{bol} \simeq 4\un{nA}$. The relevant parameters of the fit are shown in the legend. Signal amplitude has been divided by the amplification given by the electronic chain, to obtain the voltage variation across the thermistor.
}

\label{fig:55Fe}
\end{figure}

Applying a Bessel-Thomson 3dB cutoff frequency $f_B=200\un{Hz}$, we varied the bias voltage in the range $1 - 30\un{V}$, using two different load resistances: 2 and 11$\un{G\Omega}$. Acting on the polarization circuit in this way, we were able to span a large range of bias current values, namely $150 - 15000\un{pA}$. 
We restricted our analysis to the events due to $^{55}$Fe decay. An example of the Fe double peak, as it appears in the amplitude spectrum,  is shown in Fig.~\ref{fig:55Fe}. We performed a fit with a double Gaussian function, fixing the ratio of the peak positions and amplitudes to the known values from nuclear data~\cite{chu2005lund}.

\begin{figure}[t]
\centering \includegraphics[width=0.48\textwidth]{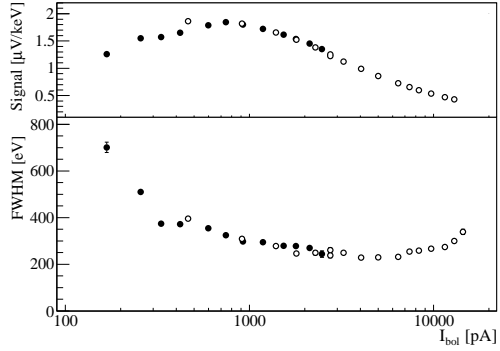}
\caption{Light detector performances resulting from the variation of the current in the bias circuit, using load resistances of $11\un{G\Omega}$ (black dots) and $2\un{G\Omega}$ (open circles). Both the signal yield (on the top) and the FWHM resolution (on the bottom) were computed on the $5.9\un{keV}$ $^{55}$Fe peak. 
}
\label{fig:amp_and_reso}
\end{figure}

From the prominent $^{55}$Fe peak we computed the signal yield as the ratio between the measured voltage variation across the thermistor and the energy of the line. The signal yield is expressed in $\mu$V/keV. We also calculated the resolution of our detector as the FWHM of the $^{55}$Fe peak in eV. Results are shown in Fig.~\ref{fig:amp_and_reso}. 
 
As we can see from the picture, the maximum of the signal yield does not coincide with the best energy resolution. This is mainly due to the non-linear behavior of the thermistor. Indeed, increasing the signal yield also enlarges the noise in a not easily predictable way.
As a consequence of the  optimum filter application, the energy resolution depends on both noise power spectrum and signal shape, representing the true expression of the signal to noise ratio. Therefore, the optimal working point of the LD should be chosen as the one corresponding to the minimum of this curve.
The value of the energy resolution of our detector at the optimal working point (that is not a single point but a large plateau between 2000 and $7000\un{pA}$) is $\sim 250\un{eV}$. 

In both the plots of Fig.~\ref{fig:amp_and_reso} we do not observe deviations from the common trends corresponding to variations of the load resistance, so we conclude that the value of the load resistance does not affect the signal amplitude and the energy resolution at the present level of accuracy. This is consistent with the a priori estimation of the load resistance Johnson noise discussed in sec.~\ref{sec:ld_operation}, since a contribution of $\sim 1-30\un{eV}$ rms is negligible with respect to the measured values of  Fig.~\ref{fig:amp_and_reso}.
The general decreasing behavior of the energy resolution is interrupted at very high polarization currents. At these values, in fact, the parallel noise contributions become smaller than the ones given by the readout circuit (in particular, by the preamplifiers) and, from this point on, noise assumes the constant value of $\sim 50\un{nV}$ rms. Simultaneously, signal yield continues to decrease, determining the energy resolution worsening reported in the plot.
It is important to stress that, however, a reduction of preamplifiers noise would not produce a sensible improvement of the resolution, since the signal yield decreases faster than the bias-related noise.

\begin{figure}[t]
\centering \includegraphics[width=0.48\textwidth]{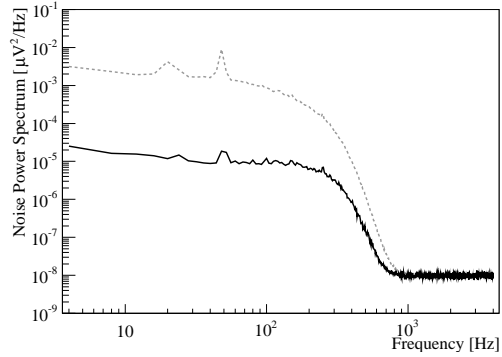}
\caption{Light detector noise power spectra for low (240 pA) and high (15000 pA) current, respectively in gray dotted line and in black solid line. Both the spectra were obtained averaging randomly triggered events including no pulses. }
\label{fig:nps}
\end{figure}
\begin{figure}[t]
\centering \includegraphics[width=0.48\textwidth]{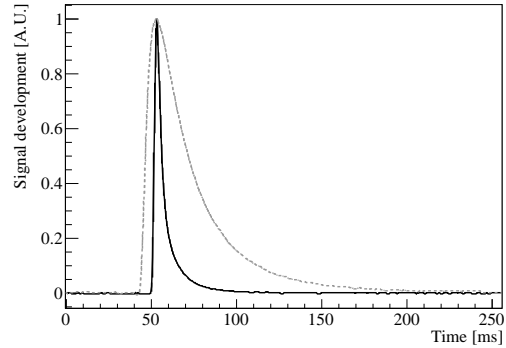}
\caption{Light detector average pulse for low (240 pA) and high (15000 pA) current, respectively in gray dotted line and in black solid line. The reported waveforms were obtained averaging pulses belonging to the $^{55}$Fe lines.}
\label{fig:avg}
\end{figure}

Figures \ref{fig:nps} and \ref{fig:avg} show the noise power spectrum and the average detector response for low (240 pA) and high (15000 pA) bias current. The noise power spectrum was computed averaging randomly triggered waveforms including no pulses. The average pulse was calculated as the mean of a large number of pulses belonging to the $^{55}$Fe lines. 
When comparing the noise power spectra we should keep in mind that the relevant part is the low frequency region ($0-100\un{Hz}$) where the signal spectrum is confined. In this region, the increase of the bias current reflects into a reduction of the noise power spectrum of about two orders of magnitude. This is consistent with the energy resolution trend commented above.

For what concerns the average detector response, the faster signal development in the case of higher bias is clear by looking at Fig.~\ref{fig:avg}. We analyzed the distributions of the rise and decay times of $^{55}$Fe pulses. 
\begin{figure}[t]
\centering \includegraphics[width=0.48\textwidth]{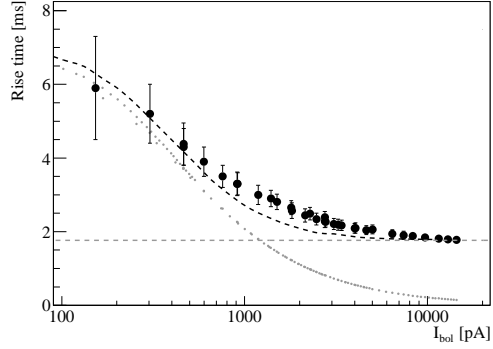}
\caption{Light detector signal rise time (black dots) as a function of the current in the bias circuit. The reported values were computed on pulses belonging to the $^{55}$Fe lines. The trend is well reproduced by the convolution (black dashed line) of two time responses: one due to the $200\un{Hz}$ Bessel-Thomson filter (gray dashed line) and the other given by the readout circuit (gray dotted line).}
\label{fig:rise1}
\end{figure}

The rise time is computed as the time difference between the $90\%$ and the $10\%$ of the leading edge. The mean value of the pulse rise time as a function of the bias current is shown in Fig.~\ref{fig:rise1}. The detector is remarkably faster ($\sim 3$ times) at high bias currents. Speed response is limited by four different components. The first is introduced by the presence of the Bessel-Thomson filter, which gives a contribution that is independent from the thermistor impedance and, consequently, from the bias current. We evaluated this contribution by convolving in the frequency domain a step function with the transfer function of our Bessel-Thomson filter. This is reported in Fig.~\ref{fig:rise1} by a dashed line.
The second component is the rise time of the readout circuit, whose elements are the thermistor with its dynamic impedance, and the readout wires, having a parasitic capacitance of $\sim 250 \un{pF}$ and a negligible resistance. While the capacitance of the readout wires is constant, the working resistance of the thermistor depends on the bias current, as explained in section~\ref{sec:ld_operation}. The rise time of the readout circuit as a function of the bias current can be seen in Fig.~\ref{fig:rise1} (gray dotted line).
The last  two contributions are given by the intrinsic thermal response of the absorber and the thermistor. They cannot be reliably estimated due to the high complexity of the phonon dynamics in the two components. 
Nevertheless the behavior of the experimental data is well reproduced by the convolution of the first two time responses (black dashed line in Fig.~\ref{fig:rise1}). Indeed, for low bias currents the RC contribution is dominant, while for higher currents the Bessel-Thomson filter accounts alone for the observed rise time. This implies that the thermal rise time, whatever it is, is negligible.
When considering the rise time, it is important to underline that having fast signals reduces the probability of pile-ups that could spoil the detection efficiency. Therefore, large bias currents improve light detector performances  in terms of both signal to noise ratio and detection efficiency.

\begin{figure}[t]
\centering \includegraphics[width=0.48\textwidth]{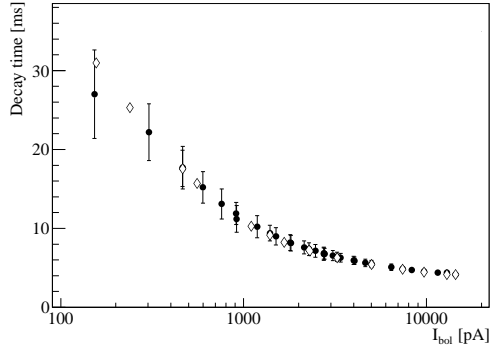}
\caption{Light detector signal decay time as a function of the current in the bias circuit (black dots). The evaluated decay times of the thermal pulses, evaluated by fitting the pulses with Eq.~\protect\ref{eq:thermalpulse} as fit function, are also shown as open diamonds. All the reported values were computed on pulses belonging to the $^{55}$Fe lines.}
\label{fig:decay1}
\end{figure}

\begin{figure}[t]
\centering \includegraphics[width=0.48\textwidth]{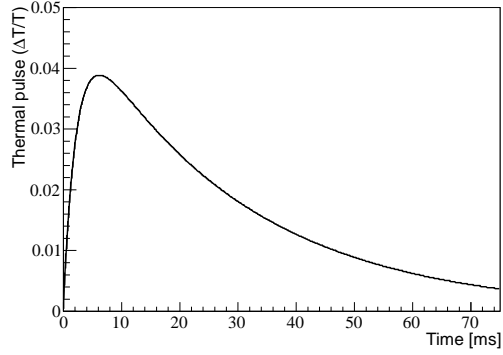}
\caption{Example of thermal pulse development for a bias current $I_{bol}=150\un{pA}$. The amplitude is expressed in the adimensional form $\frac{\Delta T(t)}{T}$ ($T$ is the LD working temperature). }
\label{fig:thermpulse}
\end{figure}

The decay time is computed as the time difference between the $30\%$ and $90\%$ of the pulse trailing edge. The mean value of the pulse decay time as a function of the bias current is shown in Fig.~\ref{fig:decay1}. As for the rise time, we can identify four different contributions to the decay time: Bessel-Thomson, readout, thermal response of the absorber and the thermistor. The first two have been already estimated for the rise time and are of the order of $2-6\un{ms}$, well below the decay time measured values. We deduce that the absorber thermal component is dominant in the decay time, provided that the thermistor contribution can be considered negligible. To confirm this hypothesis we performed a fit to the average detector response for a given bias current, in order to extract the decay time of the thermal component alone. The fit function is based on the thermal model response described in~\cite{Carrettoni:2011rn,Vignati:2010yf}. The fit accepts as input parameters measurable quantities like the sampling frequency, the gain, the Bessel-Thomson cutoff frequency, the bias voltage, the baseline voltage, the working resistance, the load resistance and the capacitance and gives as results the parameters of the thermal pulse which represents the temperature variation of the bolometer generated by an energy deposition:
\begin{equation}
\Delta T(t) = A_{T}(-e^{-\frac{t}{\tau_{r}}}+ \alpha e^{-\frac{t}{\tau_{d1}}} (1-\alpha) e^{-\frac{t}{\tau_{d2}}}) 
\label{eq:thermalpulse}
\end{equation}
Formula~\ref{eq:thermalpulse} indicates that the thermistor temperature increases with one rise time, $\tau_{r}$, and decreases with two decay constants, $\tau_{d1}$ and $\tau_{d2}$. The parameter $\alpha$ weights the two exponential decays and satisfies the condition $0\leq\alpha\leq1$. The amplitude of the thermal pulse, $A_{T}$ is also a function of the thermal elements and is directly proportional to the energy release $E$.
An example of thermal pulse can be seen in Fig.~\ref{fig:thermpulse}.

The decay time of the thermal component as a function of the bias current can be seen in Fig.~\ref{fig:decay1} (open diamonds) and, as expected, it accounts entirely for the observed behavior.

\subsection{Variation of the Bessel-Thomson cutoff frequency}
\begin{figure}[t]
\centering \includegraphics[width=0.48\textwidth]{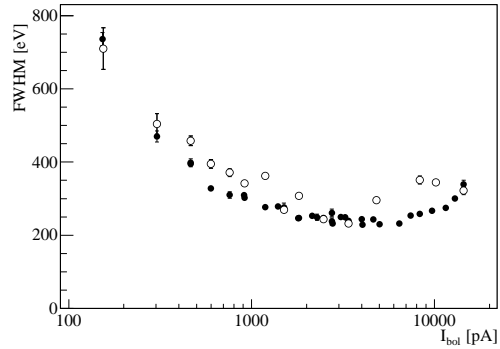}
\caption{Light detector FWHM resolution as a function of the current in the bias circuit. Data were acquired using two different Bessel-Thomson cutoff frequencies: $200\un{Hz}$ (black dots) and $550\un{Hz}$ (open circles). The reported values were computed on pulses belonging to the $^{55}$Fe lines.}
\label{fig:reso_bessel}
\end{figure}

Data were acquired using two different values of the Bessel-Thomson cutoff frequency: $200\un{Hz}$ and $550\un{Hz}$. Both the frequencies are well below the Nyquist limit of the sampling frequency ($4\un{kHz}$), avoiding aliasing phenomena. The FWHM energy resolution as a function of the bias current is shown in Fig.~\ref{fig:reso_bessel}. The $200\un{Hz}$ data are a subset of the ones already shown in Fig~\ref{fig:amp_and_reso}.

We note that an higher Bessel-Thomson cutoff frequency causes a general worsening of the energy resolution, even if, applying a bias current around 3000 pA, we obtain the same minimum value of FWHM $\simeq 250\un{eV}$ discussed above. 

\begin{figure}[htb]
\centering \includegraphics[width=0.48\textwidth]{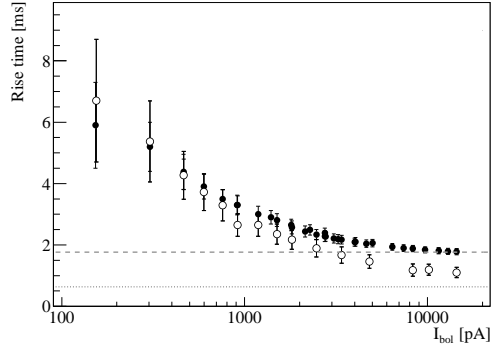}
\caption{Light detector signal rise time as a function of the current in the bias circuit. Data were acquired using two different Bessel-Thomson cutoff frequencies: $200\un{Hz}$ (black dots, same data reported in Fig.~\protect\ref{fig:rise1}) and $550\un{Hz}$ (open circles). The reported values were computed on pulses belonging to the $^{55}$Fe lines. Estimated contributions of $200\un{Hz}$(dashed line) and $550\un{Hz}$ (dotted line) Bessel-Thomson filters are also shown.}
\label{fig:rise2}
\end{figure}

For the same sets of data we have analyzed the distributions of the rise and decay times of $^{55}$Fe pulses. 
The mean values of the rise time as a function of the bias current is shown in Fig.~\ref{fig:rise2}.  As expected, a higher Bessel-Thomson cutoff frequency corresponds to a slightly faster rise time, especially at high bias currents. However, even considering the positive effect of a faster signal on the pile-up rejection, the general worsening of the energy resolution leads us to prefer lower Bessel-Thomson cutoff frequencies as the standard choice.

No effect was observed on the decay time distribution where, as explained in the previous section, the thermal component is dominant.

\subsection{Variation of the working temperature} \label{sec:temp}
\begin{figure}[t]
\centering \includegraphics[width=0.48\textwidth]{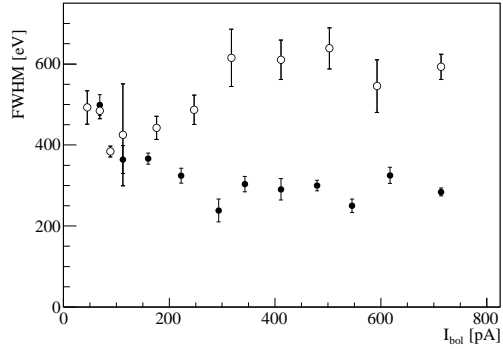}
\caption{Light detector FWHM resolution as a function of the current in the bias circuit. Data were acquired at two different cryostat temperatures: $\sim20\un{mK}$ (black dots) and $\sim10\un{mK}$ (open circles). The reported values were computed on pulses belonging to the $^{55}$Fe lines. We note that the best energy resolution achieved using this detector at $\sim20\un{mK}$ is comparable with the one of the other LD (see sec.~\protect\ref{sec:bias}).
}
\label{fig:temp}
\end{figure}

In order to investigate the effect of the working temperature on the LD performances, we carried out several sets of measurements with various Ge light detectors, including the one presented so far in this work, varying the working temperature of the cryostat from $20\un{mK}$ to a significantly lower value.
Among these measurements, we present here the one with the lowest achieved temperature ($\sim 10\un{mK}$). The light detector used in this run is very similar to the one presented above, namely a Ge disk with dimensions $\diameter 50\un{mm} \times 266\un{\mu m}$. It was mounted and operated in exactly the same way and, also, equipped with the same $^{55}$Fe source.
In Fig.~\ref{fig:temp}, the FWHM energy resolution as a function of the bias current is shown for the two different values of the cryostat working temperature: $\sim20\un{mK}$ and $\sim10\un{mK}$.  

With the exception of the low bias regime ($< 200\un{pA}$), the higher temperature measurements clearly show better performances (roughly a factor 2 on the FWHM). 
This is somehow in contradiction with what we expected. In fact, a lower temperature determines smaller heat capacities and, hence, higher signal yield. At the same time, however, the change of working temperature modifies the whole thermal circuit, thus determining possible changes in the shape of both signal and noise.
We tried to investigate the source of this behavior by looking at the average signal and noise power spectrum.
In particular, cooling down to $\sim10\un{mK}$, we observe mild alterations in the signal shape while the noise power spectrum in the lowest frequencies region ($0-50 \un{Hz}$) increases. This causes a general worsening of the energy resolution.

\section{Conclusions}
We investigated the performances of Ge bolometric light detectors in terms of signal amplitude, energy resolution and signal time development. Data show a clear dependence on the applied bias current, favoring values around $2-7\un{nA}$.  At higher polarization currents, series noise due to preamplifiers becomes dominant and limits the signal to noise ratio. 
The value of the load resistance has no influence on the signal amplitude and the energy resolution, confirming that the Johnson noise does not remarkably affect the light detector performances.
Different choices of the Bessel-Thomson filter cutoff frequency do not critically change the LD behavior, provided that, however, too low or too high values would spoil the energy resolution.
Temperature plays an important role: LD performances at $\sim 20\un{mK}$ are significantly better with respect to the ones observed at $\sim 10\un{mK}$,  the standard working temperature of bolometers for rare events searches.

\begin{acknowledgments}
The results reported here have been obtained in the framework of the Lucifer Experiment funded from the
European Research Council under the European Unions Seventh Framework Programme (FP7/2007-2013) / ERC
grant agreement n. 247115.
Thanks are due to E. Tatananni, A. Rotilio, A. Corsi, B. Romualdi and F. De Amicis for continuous and constructive help in the  overall setup construction. Finally, we are especially grateful to M. Perego and M. Guetti for their invaluable help.
\end{acknowledgments}

\bibliographystyle{JHEP} 
\bibliography{main}

\providecommand{\href}[2]{#2}\begingroup\raggedright\begin{thebibliography}{10}

\bibitem{bolo-review1}
B.~C. N.~Booth and E.~Fiorini, {\it Low temperature particle detectors},  {\em
  Annu. Rev. Nucl. Part. Sci.} {\bf 46} (1996) 471.

\bibitem{bolo-review2}
C.~Enss and D.~McCammon, {\it Physical principles of low temperature detectors:
  Ultimate performance limits and current detector capabilities},  {\em J. Low
  Temp. Phys.} {\bf 151} (2008) 5.

\bibitem{Avignone:2007fu}
F.~T. Avignone, S.~R. Elliott, and J.~Engel, {\it {Double Beta Decay, Majorana
  Neutrinos, and Neutrino Mass}},  {\em Rev. Mod. Phys.} {\bf 80} (2008) 481,
  [\href{http://xxx.lanl.gov/abs/0708.1033}{{\tt arXiv:0708.1033}}].

\bibitem{Clemenza:2011zz}
M.~Clemenza, C.~Maiano, L.~Pattavina, and E.~Previtali, {\it {Radon-induced
  surface contaminations in low background experiments}},  {\em Eur. Phys. J.
  C} {\bf 71} (2011) 1805.

\bibitem{de2003experimental}
P.~De~Marcillac {\em et.~al.}, {\it Experimental detection of
  $\alpha$-particles from the radioactive decay of natural bismuth},  {\em
  Nature} {\bf 422} (2003) 876.

\bibitem{cozzini2004detection}
C.~Cozzini {\em et.~al.}, {\it Detection of the natural $\alpha$ decay of
  tungsten},  {\em Phys. Rev. C} {\bf 70} (2004) 064606.

\bibitem{Arnaboldi:2010tt}
C.~Arnaboldi {\em et.~al.}, {\it {CdWO$_4$ scintillating bolometer for Double
  Beta Decay: Light and Heat anticorrelation, light yield and quenching
  factors}},  {\em Astropart. Phys.} {\bf 34} (2010) 143,
  [\href{http://xxx.lanl.gov/abs/1005.1239}{{\tt arXiv:1005.1239}}].

\bibitem{ZnSe2013}
J.~Beeman {\em et.~al.}, {\it Performances of a large mass {ZnSe} bolometer to
  search for rare events},  {\em submitted to JINST}
  [\href{http://xxx.lanl.gov/abs/1303.4080}{{\tt arXiv:1303.4080}}].

\bibitem{Beeman:2012gg}
J.~Beeman {\em et.~al.}, {\it {Performances of a large mass ZnMoO$_4$
  scintillating bolometer for a next generation 0vDBD experiment}},  {\em Eur.
  Phys. J. C} {\bf 72} (2012) 2142.

\bibitem{Beeman:2013Pb}
J.~Beeman {\em et.~al.}, {\it {New experimental limits on the $\alpha$ decays
  of lead isotopes}},  {\em Eur. Phys. J. A} {\bf 49} (2013), no.~4 1--7.

\bibitem{Beeman:2011yc}
J.~Beeman {\em et.~al.}, {\it {Discrimination of alpha and beta/gamma
  interactions in a TeO$_2$ bolometer}},  {\em Astropart. Phys.} {\bf 35}
  (2012) 558--562, [\href{http://xxx.lanl.gov/abs/1106.6286}{{\tt
  arXiv:1106.6286}}].

\bibitem{Itoh:1996}
K.~M. Itoh {\em et.~al.}, {\it Hopping conduction and metal-insulator
  transition in isotopically enriched neutron-transmutation-doped
  $~^{70}${G}e:{G}a},  {\em Phys. Rev. Lett.} {\bf 77} (1996) 4058.

\bibitem{Arnaboldi:2006mx}
C.~Arnaboldi, G.~Pessina, and S.~Pirro, {\it {The cold preamplifier set-up of
  CUORICINO: Towards 1000 channels}},  {\em Nucl. Instrum. Meth. A} {\bf 559}
  (2006) 826.

\bibitem{Arnaboldi:2004jj}
C.~Arnaboldi {\em et.~al.}, {\it {The front-end readout for CUORICINO, an array
  of macro-bolometers and MIBETA, an array of mu-bolometers}},  {\em Nucl.
  Instrum. Meth. A} {\bf 520} (2004) 578.

\bibitem{Gatti:1986cw}
E.~Gatti and P.~F. Manfredi, {\it Processing the signals from solid state
  detectors in elementary particle physics},  {\em Riv. Nuovo Cimento} {\bf 9}
  (1986) 1.

\bibitem{Radeka:1966}
V.~Radeka and N.~Karlovac, {\it Least-square-error amplitude measurement of
  pulse signals in presence of noise},  {\em Nucl. Instrum. Methods} {\bf 52}
  (1967) 86.

\bibitem{Arnaboldi:2010jx}
C.~Arnaboldi {\em et.~al.}, {\it {Characterization of ZnSe scintillating
  bolometers for Double Beta Decay}},  {\em Astropart. Phys.} {\bf 34} (2011)
  344, [\href{http://xxx.lanl.gov/abs/1006.2721}{{\tt arXiv:1006.2721}}].

\bibitem{chu2005lund}
S.~Y.~F. Chu, L.~P. Ekstrom, and R.~B. Firestone, {\it {The Lund/LBNL nuclear
  data search}},  {\em Available at nucleardata. nuclear. lu.
  se/nucleardata/toi}.

\bibitem{Carrettoni:2011rn}
M.~Carrettoni and M.~Vignati, {\it {Signal and noise simulation of CUORE
  bolometric detectors}},  {\em JINST} {\bf 6} (2011) P08007,
  [\href{http://xxx.lanl.gov/abs/1106.3902}{{\tt arXiv:1106.3902}}].

\bibitem{Vignati:2010yf}
M.~Vignati, {\it {Model of the Response Function of Large Mass Bolometric
  Detectors}},  {\em J. Appl. Phys.} {\bf 108} (2010) 084903,
  [\href{http://xxx.lanl.gov/abs/1006.4043}{{\tt arXiv:1006.4043}}].

\end{thebibliography}\endgroup

\end {document}